\documentclass[pre,twocolumn, showpacs, preprintnumbers, amsmath,
amssymb]{revtex4}
\usepackage[pdftex]{graphicx}
\usepackage{amssymb}
\usepackage{amsmath}
\usepackage{graphicx}
\usepackage{epsfig}
\usepackage{epstopdf}
\usepackage{dcolumn}
\usepackage{bm}

\begin{document}

\title{Soap films as two-dimensional fluids: Diffusion and flow fields}
\author{Skanda~Vivek}
\author{Eric R.~Weeks}
\affiliation{Department of Physics, Emory
University, Atlanta, GA 30322}
\date{\today}

\begin{abstract}
We observe tracer particles diffusing in soap films to measure
the two-dimensional (2D)
viscous properties of the films.  We make soap films with a
variety of water-glycerol mixtures and of differing thicknesses.
The single-particle diffusivity relates closely to parameters of
the film (such as thickness $h$) for thin films, but the relation
breaks down for thicker films.  Notably, the diffusivity is faster
than expected for thicker films, with the transition at $h/d = 5.2
\pm 0.9$ using the tracer particle diameter $d$.  This indicates
a transition from purely 2D diffusion to diffusion that is more
three-dimensional.  Additionally, we measure larger length scale
flow fields from correlated particle motions and find good agreement
with what is expected from theory of 2D fluids for all our films,
thin and thick.  We measure the effective 2D viscosity of a soap
film using single-particle diffusivity measurements in thin films,
and using the two-particle correlation measurements in all films.
\end{abstract}


\maketitle


\section{Introduction} 
\label{sec:intro}

Soap films are thin liquid films, stabilized by two
surfactant layers on either side. Soap films have complex
hydrodynamics~\cite{hydrodynamics} that have been widely
investigated as early as by Plateau~\cite{plateau} and
Gibbs~\cite{gibbs}. Previous experiments have demonstrated that
thin soap films behave in many respects as two-dimensional (2D)
fluids~\cite{vikPRE,vikPRL,singlesoapfilm,1dflags,rutgers98}.
Soap films have applications as a wide range of model
systems.  For example, soap films share similarities to cell
membranes~\cite{soaplipid}.  Soap films were used to study swimming
fish and flapping flags in a two dimensional wind~\cite{1dflags}.
Quickly flowing soap films also served as model systems for 2D
turbulence~\cite{rutgers98,rutgers98a,burgess99}, which is relevant
in our atmosphere at large scales.

We are interested in understanding soap film hydrodynamics
by placing tracer particles and analyzing their diffusive
motion~\cite{vikPRE,vikPRL,singlesoapfilm}. Particle motions in
a soap film are constrained in the third direction, due to small
film thickness. Hence, their diffusive motion is two-dimensional
and controlled by an effective 2D viscosity of the soap film,
$\eta_{2D}$.  $\eta_{2D}$ is expected to be related to the film
thickness and other details of the soap solution using the 1957
Trapeznikov approximation~\cite{trapeznikov}.  In 1975 Saffman
and Delbr{\"u}ck argued that diffusive motion in a fluid membrane
is also influenced by the surrounding three-dimensional (3D)
viscous fluids with viscosity $\eta_{3D}$ on either side of the
membrane \cite{sd}.  In the case of interest to soap films, the
surrounding 3D fluid is air with viscosity $\eta_{3D}=\eta_{air}$.
The Saffman-Delbr{\"u}ck approximation~\cite{sd,hydrodynamicssd}
relates the observable single particle diffusivity $D$
to $\eta_{2D}$, $\eta_{air}$, and the particle diameter $d$,
allowing one to determine $\eta_{2D}$ by observing tracer particle
trajectories \cite{vikPRE}.

The Saffman-Delbr{\"u}ck approximation is no longer applicable
in two limits.  For the first limit, note that
$\eta_{3D}$ has units of Pa$\cdot$s and $\eta_{2D}$ has units
of Pa$\cdot$s$\cdot$m, so the ratio of these two quantities
is a length scale, sometimes termed the Saffman length $l_S =
\eta_{2D}/2 \eta_{3D}$ \cite{stoneprize}.  The first limit,
as stated by Saffman and Delbr{\"u}ck \cite{sd}, is for
situations where the lateral size $R$ of the membrane becomes small,
$R \lesssim l_S$.  The crossover to the system size-limited behavior
has been seen experimentally by two groups \cite{eremin11,domanov2011},
and the behavior observed in this limit matches the predicted
behavior \cite{sd}.  The second limit is implicit.  The
Saffman-Delbr{\"u}ck approximation considers the diffusive motion of
a thin disk, diameter $d$ and height $h$.  The 2D fluid is modeled
as a 3D fluid of thickness $h$ and viscosity $\eta_B$, resulting
in an effective 2D viscosity $\eta_{2D} = \eta_B h$.  The implicit
limit then is that one would not expect this approximation to
be valid for the diffusive motion of small spheres of diameter $d
\ll h$.  There has only been minimal experimental
exploration of this limit \cite{vikPRE}.  In this prior work, it
was demonstrated that the Saffman-Delbr{\"u}ck approximation breaks
down for $h/d = 7 \pm 3$.  The large error bars were due to a lack
of data in the regime $4 \leq h/d \leq 10$.  For thin films $h/d <
4$, it was found that diffusive measurements interpreted with the
Saffman-Delbr{\"u}ck approximation led to results in agreement with
the prediction of Trapeznikov \cite{trapeznikov} for $\eta_{2D}$.

In this work, we present new experimental data of the diffusivity
of particles in soap films, to examine more closely the breakdown
of the cylindrical assumption of the Saffman-Delbr{\"u}ck
approximation.  We find that this approximation no longer holds
for $h/d > 5.2\pm 0.9$.  We additionally examine the correlated
motion of pairs of particles as a function of their separation to
independently infer $\eta_{2D}$, and demonstrate that the Trapeznikov
prediction is valid for all soap films, independent of $h/d$.  
Our results show that this correlated particle motion is
the most effective way to measure $\eta_{2D}$ from observing
diffusive motion of tracers in a soap film.  Further background
discussion of the Saffman-Delbr{\"u}ck and Trapeznikov
approximations is presented in Sec.~\ref{sec:th}.  Our experimental
methods are given in Sec.~\ref{sec:exp}, and our results are
provided in Sec.~\ref{sec:res}.

\section{Hydrodynamic theory}
\label{sec:th}

\subsection{Single particle diffusion in thin films}

Our starting point for diffusion is to measure the mean square
displacement of tracer particles, which is related to the
diffusion constant as
\begin{equation}
\frac{\langle \Delta r^2 \rangle}{4\tau}=D_{1p}
\label{eq:diff}
\end{equation}

\noindent Here $\tau$ is the lag time for the displacement,
$\Delta r = |\vec{r}(t+\tau) - \vec{r}(t)|$, and the subscript
$1p$ indicates that this diffusion constant $D_{1p}$ is based on
averages over single particle motion.  The factor of 4 is twice
the dimensionality of the measurements, thus 4 for our 2D soap
films and 6 for 3D measurements.  In 3D, the single particle
diffusion constant relates to the 3D viscosity $\eta_{3D}$ by the
Stokes-Einstein-Sutherland equation~\cite{einstein1905,sutherland1905}
\begin{equation}
D_B=\frac{k_BT}{3\pi \eta_B d},
\label{eq:stokesein}
\end{equation}

\noindent with Boltzmann constant $k_B$, absolute temperature $T$,
and particle diameter $d$.  However, this equation does not apply
to soap films, for two reasons.  First, as noted in
Sec.~\ref{sec:intro},
viscosity in 2D has different units:  Pa$\cdot$s$\cdot$m in 2D
as compared to Pa$\cdot$s in 3D.  Second, diffusion and flow in a
soap film is influenced by the viscosity of the surrounding air.
In 1975 Saffman and Delbr{\"u}ck treated this case, deriving an
approximation for $D_{1p}$ for the situation of a 2D membrane
with interfacial viscosity $\eta_{2D}$ with fluid of 3D viscosity
$\eta_{3D}$ on both sides of the membrane \cite{sd}.  Hughes,
Pailthorpe, and White later extended their result to higher order
in the small nondimensional parameter $\epsilon = d \eta_{3D} /
\eta_{2D}$ \cite{hpw}.  Hughes {\it et al.} derived
\begin{equation}
D_{1p}=\frac{k_BT}{4\pi \eta_{2D}}\left[ 
\ln\left( \frac{2}{\epsilon}\right)
 - \gamma_E + \frac{4}{\pi} \epsilon
- \frac{1}{2} \epsilon^2 \ln\left( \frac{2}{\epsilon}\right)
\right],
\label{eq:sd}
\end{equation}
using Euler's constant $\gamma_E=0.577$ \cite{hpw}; the first two
terms were given by Saffman and Delbr{\"u}ck \cite{sd,saffman1976}.
In particular, this derivation treated the 2D membrane as a
thin 3D layer of fluid with 3D (``bulk'') viscosity $\eta_B$,
thickness $h$, and therefore a 2D viscosity $\eta_{2D} = h \eta_B$.
They considered the diffusion of disks of diameter $d$ and height
$h$ which spanned the membrane thickness, and which only move
horizontally (within the membrane).  Equation \ref{eq:sd} works
well for small $\epsilon$ (large $\eta_{2D}$) and should be valid
up to $\epsilon \lesssim 0.6$ \cite{Langmuirmonolayer,petrov08}.
For arbitrarily large $\epsilon$ (small $\eta_{2D}$), Petrov
and Schwille \cite{petrov08,petrov12} extended Eqn.~\ref{eq:sd}
with an approximation to exact large-$\epsilon$ numerical results
of Ref.~\cite{hpw}.  While the large $\epsilon$ limit is of less
interest to small particles diffusing in soap films, we note that
their results have been experimentally verified using large tracer
sizes in lipid membranes \cite{petrov12} and liquid crystal films
\cite{sdliqfilm}.

Saffman and Delbr{\"u}ck also noted that in a small circular
membrane of radius $R$, Eqn.~\ref{eq:sd} no longer applies, but
rather a result that depends on $R$ \cite{sd}:
\begin{equation}
D_{1p}=\frac{k_BT}{4\pi \eta_{2D}}\left[ 
\ln\left( \frac{2R}{d}\right) - \frac{1}{2} \right].
\label{eq:finiter}
\end{equation}
In practice, the diffusion constant that one expects is the smaller
of the two results, Eqns.~\ref{eq:sd} and \ref{eq:finiter}.
Comparing the leading order term of these two equations shows
that the crossover is expected when $R/d \approx \epsilon^{-1}
= \eta_{2D} / d \eta_{3D}$.  Recalling the Saffman length $l_S
= \eta_{2D}/2 \eta_{3D}$, the crossover can be expressed as
occurring at $R \approx 2 l_S$.  An alternate way to consider
this is to define $\epsilon_R = d/R$, so that Eqn.~\ref{eq:sd}
applies for $\epsilon > \epsilon_R$, and Eqn.~\ref{eq:finiter}
applies for $\epsilon < \epsilon_R$.  The crossover at
small system size has been recently confirmed experimentally
\cite{eremin11,domanov2011}.  For our experiments (to be
described in Secs.~\ref{sec:exp} and \ref{sec:res}), $R \approx 1$ cm
and our largest particle size is $d=0.5$~$\mu$m, corresponding to a
maximum $\epsilon_R=5\cdot10^{-5}$.  Our experiments are conducted
in the range $1.2 \cdot 10^{-4} < \epsilon < 3 \cdot 10^{-2}$ and
so we are safely in the ``large film'' limit where
Eqn.~\ref{eq:sd} will apply.

Soap films are made from a regular fluid with added surfactant
molecules, and it is straightforward that the effective viscosity
$\eta_{2D}$ for a soap film should depend on its constituents.
This was first described in 1957 by Trapeznikov \cite{trapeznikov}.
Similar to Saffman and Delbr{\"u}ck, he noted that there should
be a contribution $h \eta_B$ from the bulk fluid used to make
the soap film.  Dimensionally, this makes sense, and it is
also physically reasonable that $\eta_{2D}$ should increase for larger $h$
or $\eta_B$.  Trapeznikov also noted that the surfactants at the
fluid-air interface should themselves act like a 2D fluid and
contribute their own 2D viscosity $\eta_{int}$, so therefore the
effective 2D viscosity of the entire soap film would be given by
\begin{equation}
\eta_{2D,T}=\eta_{B}h+2\eta_{int}.
\label{eq:trap}
\end{equation} 
This then is a prediction that $\eta_{2D}$ measured using
Eqns.~\ref{eq:diff} and \ref{eq:sd} is equal to $\eta_{2D,T}$.
This prediction was confirmed in prior experiments by Prasad and
Weeks for thin soap films with $h/d < 7\pm 3$ \cite{vikPRE,vikPRL},
but for thicker films diffusion seemed to sense the 3D nature of the
film and follow more closely Eqn.~\ref{eq:stokesein} \cite{vikPRL}.

\subsection{Two particle correlated motion in thin films}
\label{sec:twotheory}

Two-particle microrheology is an alternative analysis technique
that complements measuring single-particle diffusion via
Eqn.~\ref{eq:diff} \cite{two-pointmicrorh,two-pointmicrorhtheory}.
Conceptually, this examines correlations between the motion of each
pair of particles.  If our soap films obey 2D hydrodynamics,
two-particle correlations should obey 2D hydrodynamic
theory~\cite{2Dleonardo} in which the correlations decay as
$\ln(R)$, where \textit{R} is the separation between two particles.
This is in contrast to the situation in 3D, in which correlations
decay as $1/R$~\cite{two-pointmicrorh}.

Specifically, there are four eigenmodes
corresponding to pairwise motion in 2D. Two of these modes are
parallel motions (+) in the longitudinal direction (\textit{x})
and transverse direction ($y$). The other two are anti-parallel motions (-)
along $x$ and $y$. These four correlation functions are calculated by:
\begin{eqnarray}
D_{x\pm}(R,\tau)&=&\langle \frac{1}{2}[x_i(\tau)\pm
x_j(\tau)]^2\delta(R-R_{ij})\rangle_{i \neq j}\nonumber\\
D_{y\pm}(R,\tau)&=&\langle \frac{1}{2}[y_i(\tau)\pm y_j(\tau)]^2\delta(R-R_{ij})\rangle_{i \neq j}
\label{eq:d2p}
\end{eqnarray} 

\noindent
For a purely viscous system, much as $\langle \Delta r^2 \rangle
~ \tau$ (e.g., Eqn.~\ref{eq:diff}), these correlation functions
also will be proportional to the lag time $\tau$.

Di Leonardo \textit{et al.}~proposed a theory~\cite{2Dleonardo}
based on the two-dimensional Stokes equation.  The theory makes
several approximations: neglecting stresses from the bounding fluid
(air), neglecting the finite film size (in the lateral dimension),
and neglecting inertia.  The Oseen tensor is obtained from the
two-dimensional Stokes equation from which the four eigenvalues
corresponding to the eigenmodes given above can be solved
\cite{2Dleonardo}.  The solutions find correlations depending on
$R$ as:
\begin{eqnarray}
D_{x\pm}/\tau &=&A\pm B\ln\frac{L}{R}\nonumber\\
D_{y\pm}/\tau &=&A\pm B\left( \ln\frac{L}{R}-1\right)
\label{eq:condeig}
\end{eqnarray}
with
\begin{eqnarray}
A &=&2D_{1p} \nonumber\\
B &=&\frac{k_BT}{2\pi \eta_{2D}}.
\label{eq:ab}
\end{eqnarray}
$L$ is a length scale beyond which the approximation fails,
although it can fail for different reasons in differen
situations.  For example, similar
to the Saffman-Delbr{\"u}ck approximation, $L$ could be related
to the smaller of the system size $R$ and the Saffman length $l_S$
\cite{2Dleonardo}.

Note that in Ref.~\cite{2Dleonardo}, they assumed $\eta_{2D}
= h \eta_B$ for a soap film \cite{2Dleonardo}, but it has been
demonstrated that $\eta_{2D} = \eta_{2D,T}$ is more appropriate for
soap films \cite{vikPRE}.  As can be seen in Eqn.~\ref{eq:trap},
$\eta_{2D,T} \approx h \eta_B$ for thick films where $h$ is large,
so the distinction only matters for thin films.

In summary, measuring the correlations described in
Eqns.~\ref{eq:d2p}, fitting to Eqns.~\ref{eq:condeig}, and
interpreting the fit parameters with Eqns.~\ref{eq:ab} is another
route to measuring $\eta_{2D}$.  One advantage of this method is
that it should be less sensitive to the exact position of small
tracer particles within the film:  partially protruding into
the air, or fully immersed in the film.  Protrusion of a tracer
particle into the air certainly affects its single-particle
mobility \cite{spheredrag}, and so using single-particle
analysis methods may result in errors in determining $\eta_{2D}$.
In contrast, the two-particle correlations are measuring long-range
hydrodynamic correlations which are insensitive to the local details
\cite{two-pointmicrorh}.  Even if one examines correlations between
one particle protruding through the soap film surface and a second
particle fully immersed in the film, the particular motions due
to the local environment of each particle will be uncorrelated,
and the long-range correlations should feel only the hydrodynamic
effects of the soap film itself (perhaps as modified due to coupling
with the air).  Also, the two-particle correlation predictions
(Eqns.~\ref{eq:condeig}) do not make any assumptions about the
nature of the tracer particles, but focus only on hydrodynamics.
In other words, these predictions do not assume that the tracers
are embedded cylinders, unlike the Saffman-Delbr{\"u}ck approach.
So, these predictions should hold even in the limit of small tracer
particles with diameters smaller than the film thickness, $d \ll h$.
Historically, this insensitivity to the tracer details was a key
strength and motivation for two-particle correlation techniques
in soft matter \cite{two-pointmicrorh}.

\section{Experimental Methods}
\label{sec:exp}

\subsection{Samples and data acquisition}

We make our soap films from bulk solutions of water, glycerol,
and surfactant.  We use the dishwashing detergent Dawn$^{\rm TM}$
as our surfactant to stabilize the interfaces of the soap film.
Once the bulk solution is prepared, we add fluorescent polystyrene
particles of certain diameter (we use \textit{d}=0.1, 0.18, and
0.5 $\mu$m). We then draw a soap film from the bulk solution using
a rectangular metal wire frame with dimensions $\approx 1.5$~cm$\times
2.0$~cm.

We have a microscope chamber made with a water filled sponge
lining to increase humidity and reduce evaporation of water from
the soap film.  This chamber is mounted on our upright microscope,
and the freshly drawn soap film is placed inside the chamber. We
seal the chamber as far as possible from outside air, to reduce
convection at the soap-film interface.

We use fluorescence microscopy to record a 33~s movie of
particles diffusing in the soap film at a frame rate of 30
images/s. The film is illuminated using an arc lamp, and a movie
of particles diffusing is taken using a CCD camera. Microscope
objectives 20$\times$, 40$\times$, and 63$\times$ are used
for particles of diameter \textit{d}=0.5, 0.18, and 0.1 $\mu$m
respectively.  We post-process the movies using particle tracking
algorithms~\cite{idlref} to extract particle positions from
individual frames.

Some macroscopic flow of the soap film in its frame is unavoidable,
resulting in coherent flow of the tracers in our movies.  Between
each video frame we compute the center of mass motion by finding the
average displacement of every particle.  The uniform flow is then
subtracted from the particle positions to provide their relative
locations in the frame of reference co-moving with the flow.
This lets us then study the diffusive motion of the individual
particles.  However, drift removal may artificially reduce true
long-range hydrodynamic correlations.  We have checked that our
algorithm does not unduly affect the correlations; details of this
are given in the Appendix.

We would like to know where our particles are within the soap film,
but this is difficult to determine directly given that the depth of
focus of our microscope is comparable to the soap film thickness.
For particles in films thinner than the particle diameter, it
is highly likely that capillary forces ensure that the particle
lies symmetrically within the film \cite{dileonardo08}.  However,
one experiment demonstrated that pinning of the contact line at a
rough particle surface can sometimes delay reaching the equilibrium
position for time scales longer than our experiments \cite{kaz12},
and we cannot rule out that our particle positions may not be
equilibrated.  For particles in films thicker than their diameter,
particles might sit at the air-water interface to reduce the
air-water surface energy.  However, as mentioned in a prior study
of soap films \cite{vikPRL} and as we observe (Sec.~\ref{sec:res}),
small particles in very thick films diffuse as if they are in a
bulk solution of the soap film liquid.  This would not happen for
particles trapped at an interface \cite{spheredrag}.  Note that
the two-particle correlation functions should be less sensitive
to the exact positions of the tracer particles, as discussed in
Sec.~\ref{sec:twotheory}.

\subsection{Measuring soap film thickness}

After taking the movie, we take the film out of the microscope
humidity chamber and measure the film's thickness using the
infrared (IR) absorption of the water based soap
films at wavelength $\lambda=3.0$~$\mu$m.  This is based on
a previously established technique~\cite{sfinfrared} which we
briefly summarize here.  Light is incident on the soap film from
an incandescent lamp. The light passes through an optical chopper,
to chop the light at a particular frequency.  This light is then
focused on the soap film by an IR lens.  An IR filter
($3.00 \pm 0.01$~$\mu$m JML Optical Industries, LLC) in the beam
path allows only wavelength of 3.0 $\mu$m to pass through. Finally,
the light is refocused on an InAs photodiode detector (Teledyne
Judson, model J12TE2-66D-R01M) by a second IR lens. The signal from the photodetector
is obtained from a lock-in amplifier (Signal recovery, model 7265)
locking with the external reference frequency of the chopper,
which reduces noise.  We separately measure the refractive index and
absorption coefficient of each bulk solution at the same wavelength.
From measured transmittance through the film and these details of
the bulk solution, we calculate the film thickness
using a modified Beer-Lambert law that takes into account the
multiple reflections in the soap film.  This method is slightly
different from prior work \cite{vikPRE}, and is a notable
improvement in that the thickness measurement is done physically
adjacent to the microscope and thus is done within 30~s of taking
the microsocpy data, allowing for higher accuracy.

\begin{figure}[tb]
\includegraphics[width=\columnwidth]{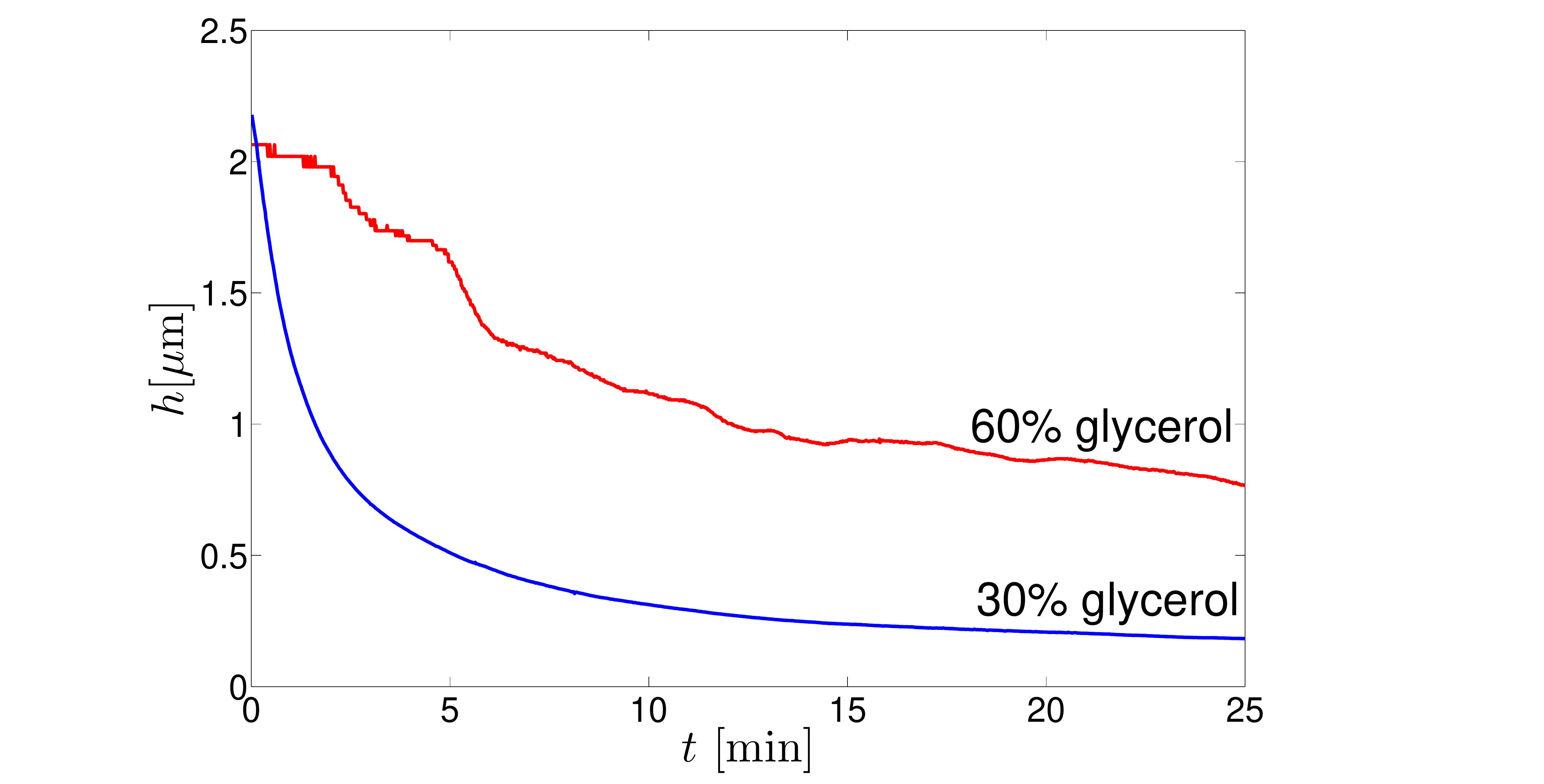}
\caption{(color online) Thickness as a function of time, which
decreases
due to soap film drainage, for films
with weight percent glycerol as indicated.
}
\label{fig:thickvstime}
\end{figure}

Soap films thin over time at the center due to drainage of
liquid towards the sides arising from capillary forces. Figure
\ref{fig:thickvstime} shows soap film drainage for 30 percent
and 60 percent glycerol weight content soap films. In general,
we observe that films made of bulk solutions with lower glycerol
concentration drain faster than those made of higher glycerol
concentration.

The timing of our experiment matters.  When the soap film is
initially drawn on to our frame, in addition to the drainage,
there are also transient flows primarily due to air currents.
Accordingly, after placing the film into the microscope chamber and
sealing the chamber, we wait for 10 to 20 minutes before taking
the movie.  This allows the initial rapid drainage to slow, and
also the transient flows.  The duration of the movie (33~s) and
delay before measuring the thickness (30~s) are short on
the time scale of the drainage, as is apparent from
Fig.~\ref{fig:thickvstime}.

In our soap film thickness measurement, there are three main sources
of error. The first source of error is due to soap film drainage in the
time between our microscopy and soap film thickness measurements.
This error is higher in low viscous soap films
due to their faster drainage as mentioned above. We quantify
this error by measuring the maximum change in thickness during
$30$~s from the curves shown in Fig.~\ref{fig:thickvstime},
which is a maximum of 0.03 $\mu$m. The second source of error
comes from the precision of the lock-in amplifier and noise
present in the measurement.  This is worse for thicker films
(which have less transmitted light) and is at most 0.02~$\mu$m.
The third source of error is from residual soap film
flows and the fact that soap films are not uniformly thick. This
error is quantified by the fluctuations in the soap film thickness
after long times, when film thickness reaches equilibrium, i.e.,
when thickness is almost constant. This error is negligible for low
viscosity films, but plays an effect in films of higher viscosity,
the maximum error being 0.02 $\mu$m.  We assume these errors are
additive, so for thick films, our thickness measurement is $\pm
0.05$~$\mu$m.

\section{Results}
\label{sec:res}

\subsection{Single particle diffusion}
\label{Subsec:onepart}

\begin{figure}[tb]
\centering
\includegraphics[width=8cm]{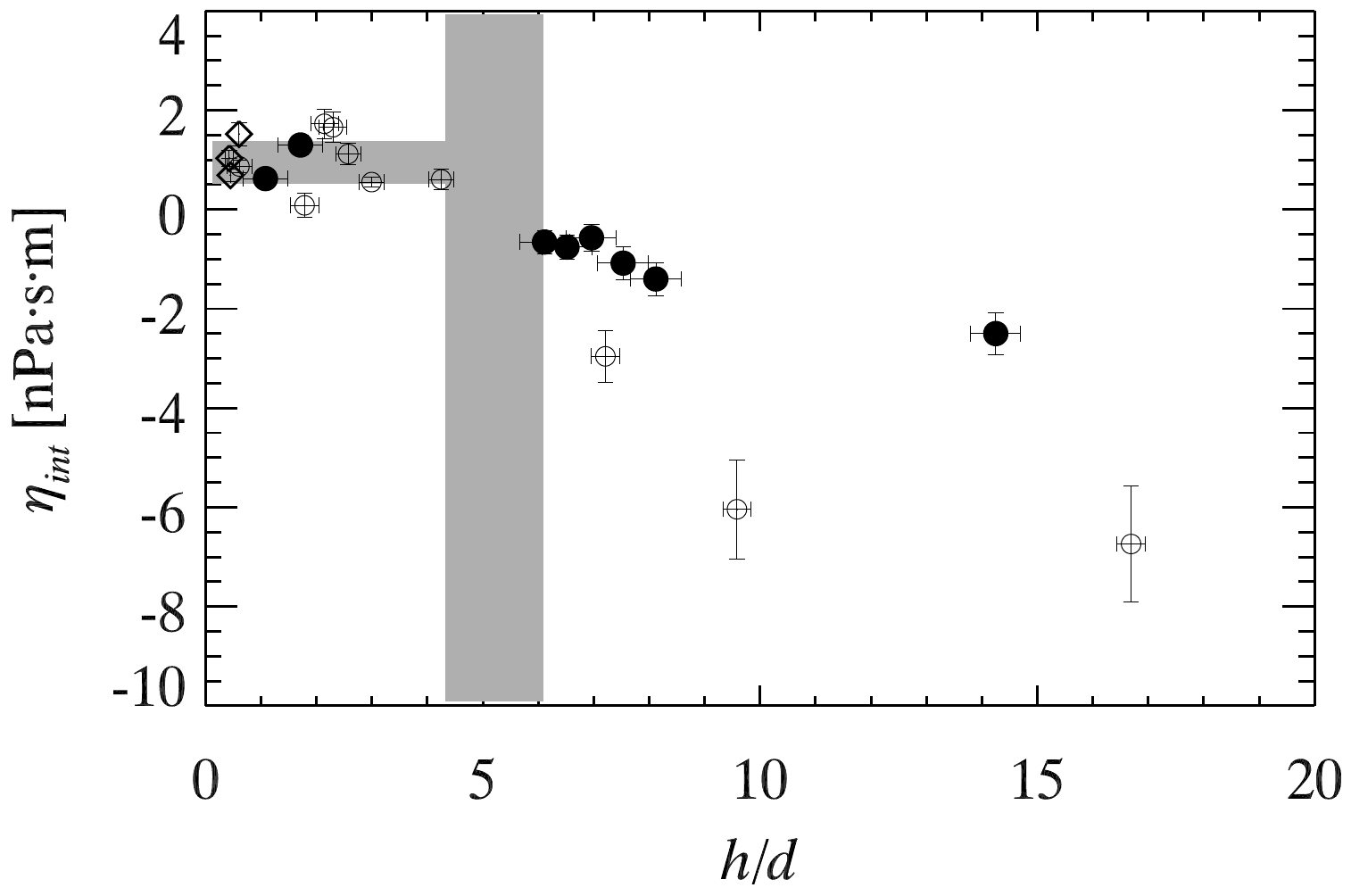}
\caption 
{Plot of interfacial viscosity from single particle diffusion measurements as a function of \textit{h/d}.
Filled circles denote particles of diameter 0.1 $\mu$m, open
circles denote particles of diameter 0.18 $\mu$m and diamonds
denote particles of diameter 0.5 $\mu$m.  The horizontal shaded
region represents $\eta_{int}=0.98 \pm 0.48$~nPa$\cdot$s$\cdot$m
based on the mean and standard deviation of the data for $h/d <
5$.  The vertical shaded region represents the transition from
physical behavior at small $h/d$ to unphysical behavior at $h/d >
5.2 \pm 0.9$.
The horizontal error bars are due to uncertainties of $h$, and
vertical error bars are due to uncertainties of $h$ and
$\eta_{2D}$ (see Eqn.~\ref{eq:trap}).
} \label{fig:nint}
\end{figure}

In a bulk (3D) liquid, diffusivity $D_B$ is a constant given
by Eqn.~\ref{eq:stokesein} \cite{einstein1905,sutherland1905}.
However, in soap films we expect diffusivity $D_{1p}$ to follow
Eqn.~\ref{eq:sd} using $\eta_{2D}$ equal to the Trapeznikov
viscosity, $\eta_{2D,T}$ given by Eqn.~\ref{eq:trap}.  Reversing
this logic, we compute $\eta_{2D}$ from measured single particle
diffusivity $D_{1p}$ using Eqn.~\ref{eq:sd}, the known viscosity
of air ($\eta_{air}$ =  0.017 mPa$\cdot $s), and the known tracer
diameter $d$.  We then obtain $\eta_{int}$ from this measured $\eta_{2D}$
via Eqn.~\ref{eq:trap}.  This interfacial viscosity should be
independent of film thickness as it is a property solely of the
soap-air interface, and the soap concentration is kept constant
throughout our experiments.  This conjectured independence is a
test of the approximations, and accordingly in Fig.~\ref{fig:nint}
we show $\eta_{int}$ as a function of \textit{h/d}.  Each data
point corresponds to a particular soap film.  
For $h/d < 5$, $\eta_{int}$ is positive and roughly
constant; in this region, the approximations of the
Saffman-Delbr{\"u}ck model work well.  Taking the mean value of
the data for $h/d < 5$ gives us 
$\eta_{int} = 0.98 \pm 0.48$~n$\cdot$Pa$\cdot$s$\cdot$m.
This agrees with a previously published value of $0.97 \pm
0.55$~n$\cdot$Pa$\cdot$s$\cdot$m for soap films made with the
same surfactant \cite{vikPRL}.
While we do not have a direct method to measure
viscosity of the soap-air interface, the rough agreement of
the measurements for $h/d < 5$ seen in Fig.~\ref{fig:nint}
demonstrate that single-particle diffusivity is one method to
measure $\eta_{int}$ for a soap film, as has been argued
previously \cite{vikPRL}.

Figure \ref{fig:nint} also shows that for larger $h/d$, $\eta_{int}$
is negative and quite variable.  The transition occurs at $h/d =
5.2 \pm 0.9$.  This value is obtained
from the gap in our data in the transition region, i.e. $5.2$
denotes the center of the gap with a width of $0.9$ on either side.
Our current observation is an improvement over prior experiments
which had a larger gap and identified the transition as $h/d =
7 \pm 3$ \cite{vikPRE}.  Furthermore, the reasonable agreement
for this transition location between the different particle sizes
(different symbols in Fig.~\ref{fig:nint}) is good evidence
that the transition is indeed a function of $h/d$.  As noted
in Sec.~\ref{sec:th}, a breakdown for large $h/d$ is expected.
The Saffman-Delbr{\"u}ck approximation treats the tracer as a
cylinder with height equal to the film thickness, which is 
dubious for $h/d > 1$.  Given that, it is remarkable that this
approximation holds up to $h/d = 5.2 \pm 0.9$.

For larger $h/d$, the interfacial viscosities in
Fig.~\ref{fig:nint} are unphysically negative, showing
that particles are diffusing faster than expected -- that
is, faster than one expects, if $\eta_{2D}$ were equal to
$\eta_{2D,T}$.  We also observe that particles of smaller
diameter are less negative in Fig.~\ref{fig:nint}.  This is
because for large $h/d$, particles diffuse more like in bulk
\cite{diffusionliqfillm}, i.e., measured diffusion $D_{1p} \approx
D_B$, as in Eqn.~\ref{eq:stokesein}. Equating this $D_{1p}$ to the
Saffman-Delbr{\"u}ck equation Eqn.~\ref{eq:sd} and approximating
the $\ln$ term as a constant, we get for large $h/d$ that the
measured $\eta_{2D}\sim\eta_B d$.  Using Eqn.~\ref{eq:trap} to
extract $\eta_{int}$ from these apparent $\eta_{2D}$ values, we
see that the particular negative values for $\eta_{int}$ will be
smaller in magnitude for smaller $d$, and that the specific values
will also depend on $\eta_B$ (which differs from film to film in
our experiments).  The differing $d$ and $\eta_B$ give rise to
the scatter of the data seen in Fig.~\ref{fig:nint} for $h/d > 5.2$.
Despite the scatter, it is apparent that the transition from the
regime where the Saffman-Delbr{\"u}ck approximation works to
where it fails is a fairly smooth transition.

\subsection{Two particle correlated motion}
\label{Subsec:twopart}

\begin{figure}[tb]
\centering
\includegraphics[width=\columnwidth]{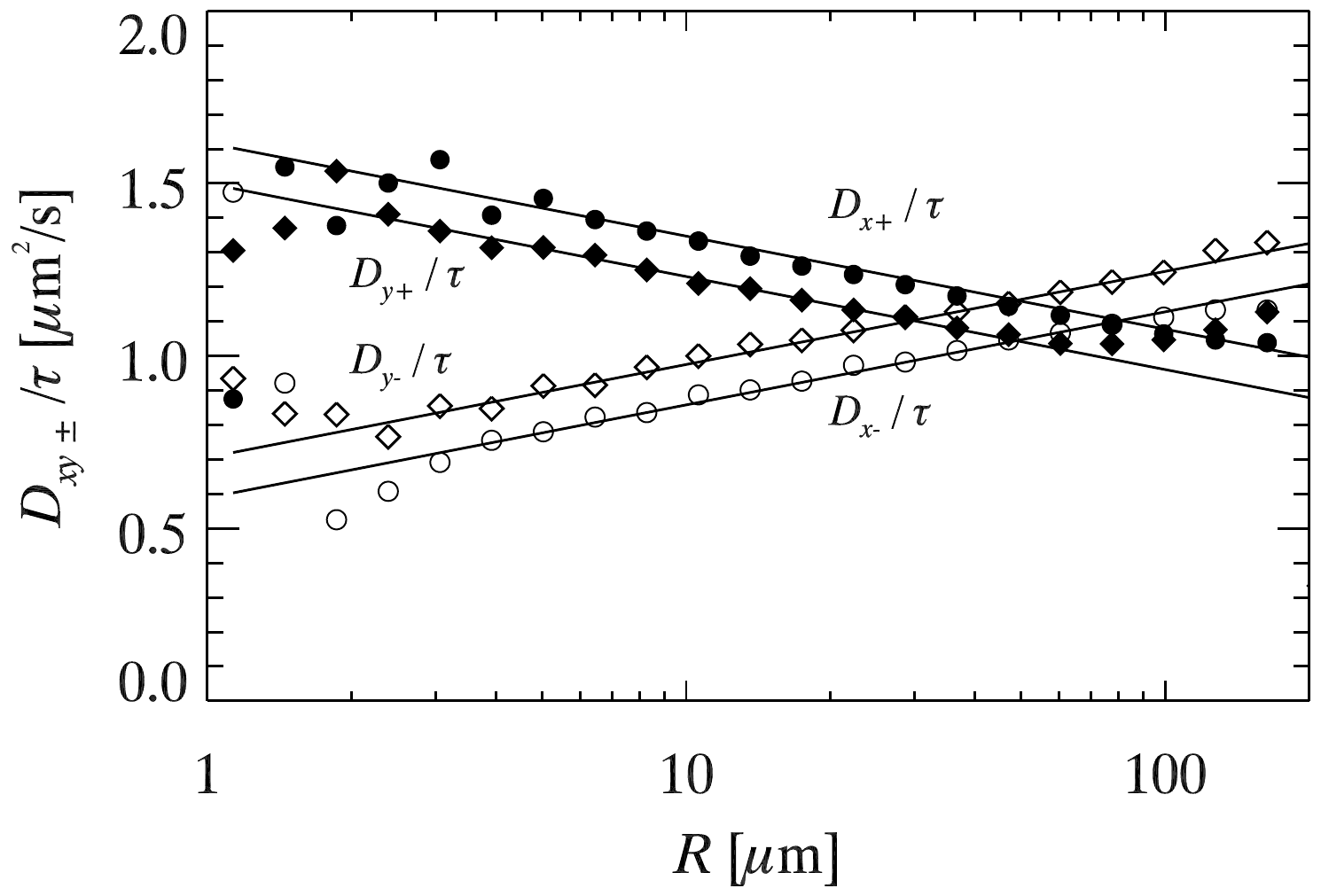}
\caption 
{Two particle correlations in a single soap film measurement as
a function of particle separation $R$. The solid lines are fits
from Eqn.~\ref{eq:condeig} with $A=1.09$ $\mu \textrm{m}^2/s$,
$B=0.12$ $\mu \textrm{m}^2/s$ and $L=81$ $\mu$m.  The data are
computed from all particle pairs and averaging over a wide range of lag times
$\tau$(see Sec.~\ref{sec:smoo}).
} \label{fig:Dxy}
\end{figure}

The two-particle measurements should not suffer from the
difficulties the one-particle measurements have, as the
two-particle correlations reveal the long-range hydrodynamic
correlations of a soap film rather than the diffusive
properties of a single particle.  As described by Eqns.~\ref{eq:d2p},
we compute the two-particle correlation functions and plot them
in Fig.~\ref{fig:Dxy} for a specific $\tau$.  The data behave as
expected.  For example, for nearby particles (small $R$), particles
move in similar directions and the parallel correlations are large
($D_{x+}$ and $D_{y+}$, indicated
by the solid symbols).  The antiparallel motions are smaller for
small $R$ ($D_{x-}$ and $D_{y-}$,
indicated by the open symbols).  All of the correlation functions
vary logarithmically with $R$, and Eqns.~\ref{eq:condeig} fit the
data well as seen by the lines.  These lines have three fitting
parameters, $A$, $B$, and $L$, which have a simple graphical
interpretation. $A$ denotes the point of intersection of positive
and negative correlations on the vertical axis, $B$ is the slope
of the lines, and $L$ is the point of intersection of $D_{x+}$
and $D_{x-}$ on the $R$-axis.

\begin{figure}[tb]
\centering
\includegraphics[width=\columnwidth]{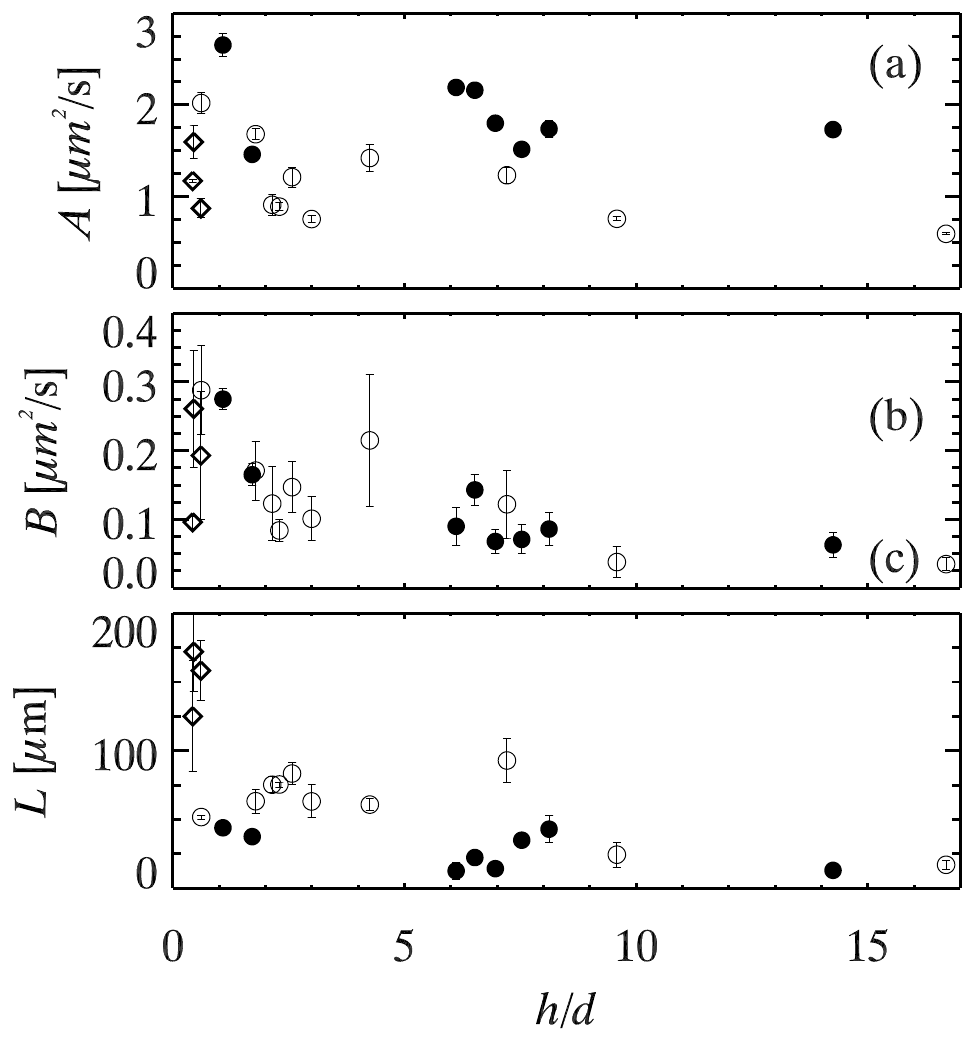}
\caption 
{Fit parameters for all experiments
as a function of $h/d$. Symbols denote particle diameters
as in Fig.~\ref{fig:nint}.  See Eqns.~\ref{eq:condeig} and
\ref{eq:ab} for the meaning of the fit parameters.  The vertical
error bars are from the standard deviations of each fit parameter
calculated for the different $\tau$'s.
} \label{fig:ABLraw}
\end{figure}

\begin{figure}[tb]
\centering
\includegraphics[width=\columnwidth]{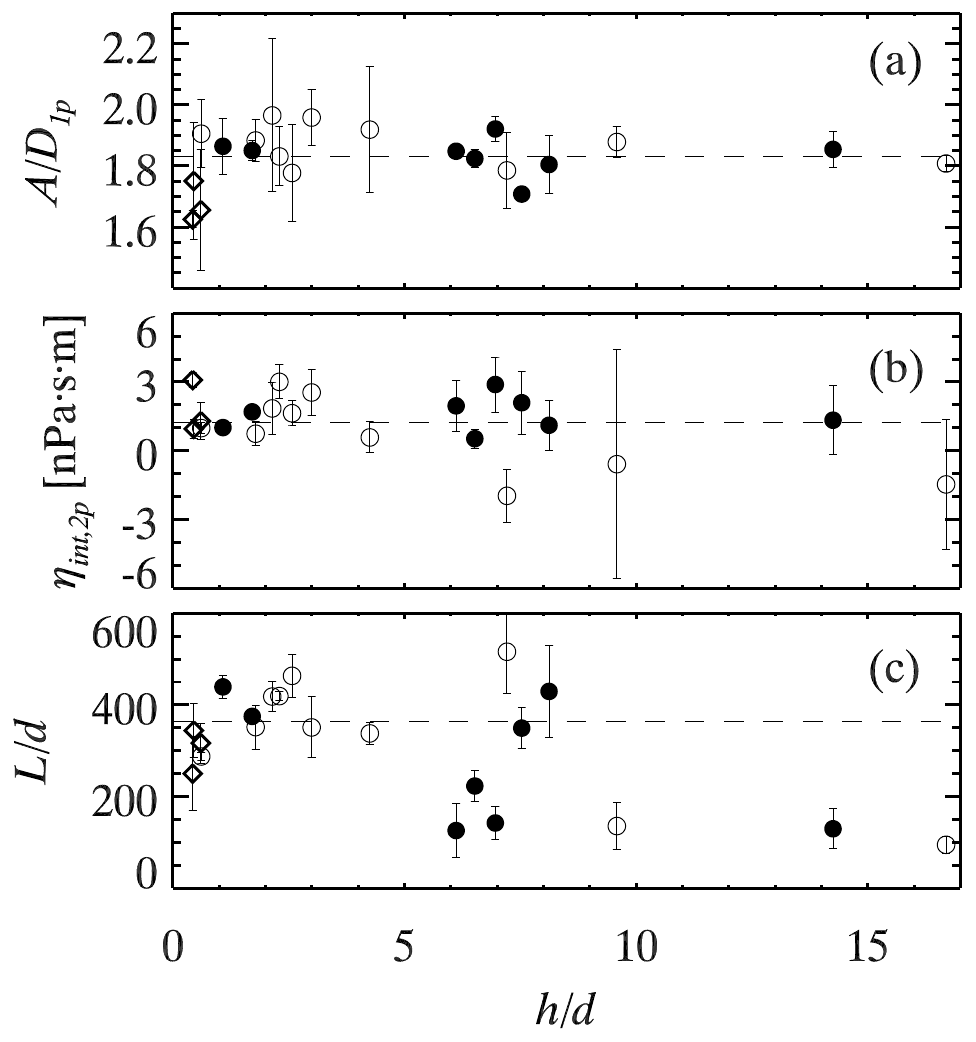}
\caption 
{Fit parameters for all experiments
as a function of $h/d$. Symbols denote particle diameters
as in Fig.~\ref{fig:nint}. (a) $A$/2$D_{1p}$ is nearly
constant, with mean value 1.83 $\pm$ 0.09 over all data as
indicated by the dashed line. (b)
$\frac{1}{2}[\frac{k_BT}{2\pi B}-\eta_Bh]$, which should
theoretically be $\eta_{int}$.  The dashed line shows
the mean value $\eta_{int,2p}$ = $1.20 \pm
1.30$ nPa$\cdot$s$\cdot$m. (c) $L/d$, where the dashed line represents
the mean value $L/d = 360 \pm 60$ for the data with $h/d < 5.2$.
} \label{fig:ABLvshd4}
\end{figure}

For each experimental movie, we compute Eqns.~\ref{eq:d2p} as a
function of $\tau$, and fit data for each $\tau$ to determine
coefficients $A(\tau), B(\tau)$, and $L(\tau)$.  As expected,
these do not depend systematically on $\tau$, and so we compute
the $\tau$-averaged values which we refer to as $A$, $B$, and $L$
for the remainder of this paper.  We plot these fit parameters
in Fig.~\ref{fig:ABLraw} as a function of $h/d$.  The data are
scattered, which is to be expected as the parameters depend on far
more than $h/d$.  The different data correspond to a variety of bulk
viscosities and particle sizes.  Accordingly, we rescale each of
these to make sense of their behavior, and show the rescaled results
in Fig.~\ref{fig:ABLvshd4}.  We now discuss these rescaled results.

From Eqn.~\ref{eq:ab}, we expect $A=2D_{1p}$ where $D_{1p}$
is the single particle diffusivity of that measurement. Figure
\ref{fig:ABLvshd4}(a) shows $A/D_{1p}$ is constant with value
$1.83 \pm 0.09$. This deviates from the predicted value,
and is consistent with previous work by Prasad and Weeks who
found a value of $1.9 \pm 0.1$.  We conjecture this deviation may be due
to the difficulty of distinguishing meaningful large $R$ parallel
correlations from correlations that are simply due to macroscopic
flow of the soap film carrying all particles in the film past our
microscope's field of view.  This does not affect the antiparallel
correlations (where any collective drift is canceled out), so
difficulties with the parallel correlation will change the points
of intersection in Fig.~\ref{fig:Dxy} that determine $A$.  This is
discussed in detail in Appendix, and the uncertainties due to this
are reflected by the error bars shown in Fig.~\ref{fig:ABLvshd4}.

Figure \ref{fig:ABLvshd4}(b) shows $\frac{1}{2}[\frac{k_BT}{2\pi
B}-\eta_Bh]$.  This quantity should be the interfacial viscosity
$\eta_{int}$ as seen by rearranging Eqn.~\ref{eq:ab}. Hence, this
is a method to find the interfacial viscosity from two particle
correlations.  Averaging all the data in Fig.~\ref{fig:ABLvshd4}(b),
we find $\eta_{int,2p}$ = $1.20 \pm 1.30$ nPa$\cdot$s$\cdot$m.  This
is consistent with our single particle measurement ($\eta_{int,1p}
= 0.98 \pm 0.48 $ nPa$\cdot$s$\cdot$m).  While the two-particle
measurement has a larger uncertainty, we believe the two-particle
value to be more reliable as it uses data from all soap films,
both thick and thin.  Moreover, this two-particle measurement is
robust to concerns about the particle position within the film,
as discussed in Sec.~\ref{sec:th}.  Our value of $A$ should
depend on the tracer details as it should be tied to $D_{1p}$
through Eqn.~\ref{eq:ab}, but $B$ and thus $\eta_{int}$ should be
measuring true properties of the soap film.


Interestingly, rescaling the third fit parameter $L$ by the particle
diameter $d$ plausibly collapses the data, especially for small
$h/d$, as shown in Fig.~\ref{fig:ABLvshd4}(c).  For thin films
($h/d < 5)$, $L/d=360 \pm 60$, indicated by the dashed line.
For thicker films $L/d$ shows scatter and for the most part
is smaller than the thin film value.  

Di Leonardo \textit{et al.}~\cite{2Dleonardo} discuss the
possible origins of the length scale $L$. In a purely theoretical
infinite-extent planar 2D fluid, there is no cutoff length scale
$L$, and correlations die out at infinity.  In reality, the finite
system size provides one potential cutoff length scale, which was
the case in their work with small films.  Our film boundary is at
least 500~$\mu$m away from our field of view, so this seems unlikely
to explain our values of $L$ of the order of 10-200~$\mu$m.
Particle motion relative to the film can lead to another length
scale
\cite{stoneprize,2Dleonardo,hydrodynamicssd}, but our particles
are passive tracers (in contrast to Ref.~\cite{2Dleonardo} for
example, which used laser tweezers to move particles).  Another
possibility is stresses from the surrounding air 
\cite{stoneprize,2Dleonardo,hydrodynamicssd}, which cannot
be neglected at distances larger than the
Saffman length $l_s$. In our system, $l_s=\eta_B h/2\eta_{air}$,
which varies from $10-1000 \mu$m. However, our observed $L$ does not
have such a wide range.  Furthermore, our thicker films generally
have higher $\eta_B$ than thinner films, and hence larger $l_s$,
yet have smaller values of $L$.  None of these length scales
seem to match our observed $L$, and these possibilities do not
explain our observed dependence of $L$ on particle size $d$.
Another possible candidate is the capillary interactions between
particles.  Previous studies found that capillary interactions
between particles in a freely suspended liquid film can cause
particle-particle interactions even at distances greater than
$100\mu$m~\cite{dileonardo08,capillarycolloidfilm,capillaryinterface},
and these interactions should scale with $d$.  These forces would
depend on if particles penetrate zero, one, or two of the film-air
interfaces, but capillary forces should not otherwise depend on the
film thickness, so the variability seen in Figs.~\ref{fig:ABLraw}(c)
and \ref{fig:ABLvshd4}(c) as $h/d$ changes seems to contradict this.
Additionally, as explained in Sec.~\ref{sec:exp}, in thick films
we think our particles are likely to be in the film interior
-- not penetrating either film-air interface -- and thus not
experiencing capillary forces.  One final possibility is that the
theory~\cite{2Dleonardo} takes only two particles into account.
We typically observe $O(50)$ particles in a field of view, and
perhaps many body effects are present in our data.  These might
affect $L$ by screening particle-particle correlations. These
effects however are more complicated to model, and determining a
length scale due to many-body effects is not possible.  Moreover,
our data are from a variety of concentrations all in the
fairly dilute limit, and concentration variations seem not
to explain the behavior of $L$.  For details, supplemental
material contains a table with all our data including
concentrations and fit parameters.  [Citation to supplemental
material will be added when
published.  A copy for review is uploaded to the manuscript
submission website, and also available at the authors' website at
http://www.physics.emory.edu/~weeks/data/ .]


\section{Conclusion} 
\label{sec:con}

We have used two different methods for measuring the effective
two-dimensional viscosity $\eta_{2D}$ of a soap film.  The 1957
paper by Trapeznikov \cite{trapeznikov} put forth Eqn.~\ref{eq:trap}
conjecturing that this viscosity is related to the soap film
thickness, the viscosity of the fluid used to form the film,
and a contribution from the surfactant layers bounding the film;
in other words, $\eta_{2D} = \eta_T$.  As we have used the same
surfactant concentration for all of our soap films, the
validation of our methods for measuring $\eta_{2D}$ is the
consistency between different measurements of $\eta_{int}$, the
contribution to $\eta_{2D}$ from the surfactant layers.  Figure
\ref{fig:nint} shows that for single-particle measurements, we
can get plausible values of $\eta_{int}$ for thin films only.
Figure \ref{fig:ABLvshd4}(b) shows that using two-particle
correlations, we get moderately consistent values of $\eta_{int}$
from all of our measurements.  The scatter of the data in both of
these figures shows that neither of these methods are fool-proof,
and best results are obtained by averaging over many films.  On
the other hand, given the variability of our tracer particle size
(a factor of 5), bulk viscosity of the soap film solutions (a
factor of 4), and film thicknesses $h$ (a factor of 30), the
agreement of the $\eta_{int}$ data is encouraging.  Our
measurement of $\eta_{int} = 1.20 \pm 1.30$ nPa$\cdot$s$\cdot$m
based on the two-particle correlations is the value we have the
most confidence in, as it uses data from every experiment we have
done and is least dependent on the details of the tracer
particles.

For larger soap films, the one-particle data of Fig.~\ref{fig:nint}
show unphysically negative $\eta_{int}$ values, whereas for the
two-particle results the data are generally physically plausible
[Fig.~\ref{fig:ABLvshd4}(b)].  The one-particle data are due to the
breakdown of the assumptions behind the Saffman-Delbr{\"u}ck
model, which models the tracers as cylinders which span the soap
film thickness.  Given this, it is pleasantly surprising that the
Saffman-Delbr{\"u}ck approach works for films up to four
times thicker than the spherical particle diameter.  The
two-particle method, in contrast, does not depend on the details
of the tracers as sensitively, but rather on the long-range
hydrodynamic properties of the soap film mas a two-dimensional
fluid.  That these hydrodynamic properties indeed behave in
a two-dimensional manner is demonstrated in Fig.~\ref{fig:Dxy} where
the two-dimensional theory fits through the data.

These methods for measuring $\eta_{2D}$ and $\eta_{int}$ should be
useful for measuring the shear viscosities of other surfactants.
Our confirmation that the flow fields are two-dimensional in
character on length scales of $5 - 100$~$\mu$m are a useful
complement to prior macroscopic experiments that treated soap
films as two-dimensional fluids
\cite{1dflags,rutgers98,rutgers98a,burgess99}.
In summary, the diffusive motion of particles appears
quasi-two-dimensional for thin films but not for thick films,
whereas the long-range flow fields appear quasi-two-dimensional
for both thick and thin films.

This work was supported by the National Science Foundation under
Grant No. CBET-1336401.  We thank V. Prasad and A. Souslov for
helpful discussions, and N. Ando, C. MacBeth, and O. Villanueva
for assistance with the infrared absorbance measurements.

\appendix
\section{Effect of drift subtraction on two particle correlation results}
\label{sec:smoo}

\begin{figure}[tb]
\includegraphics[width=\columnwidth]{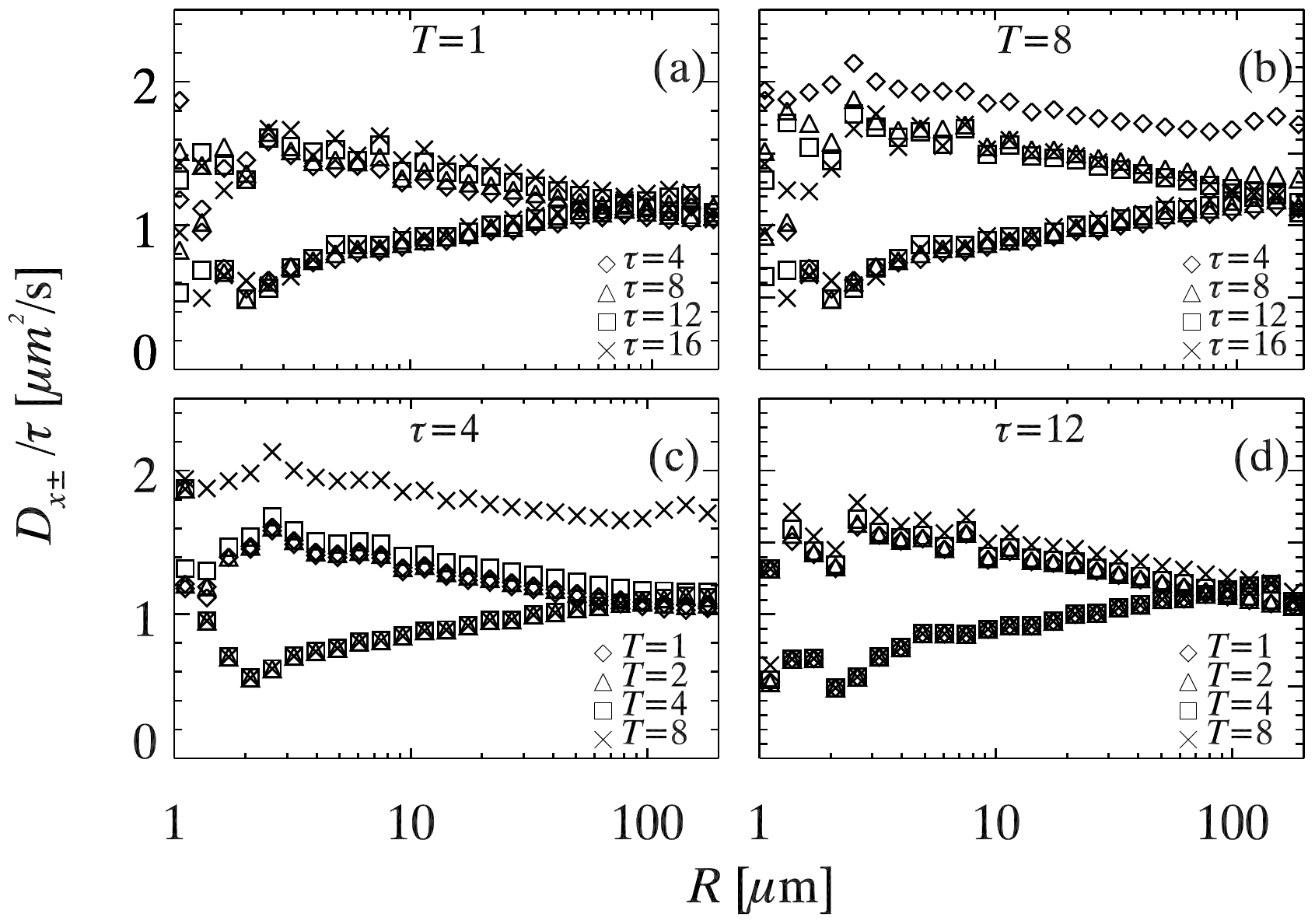}
\caption{(a) shows $D_{x+}$ and $D_{x-}$ for $T$=1. (b) shows the
same as (a) for $T$ = 8. (c) and (d) show $D_{x+}$ and $D_{x-}$
for a particular $\tau=4, 12$ respectively. Note that $T$ and
$\tau$ are in units of video frames, where 1 frame = $1/30$s. }%
\label{fig:smoo}
\end{figure}

As mentioned in Sec.~\ref{sec:exp}, some macroscopic flow of the
soap film within the soap film holder is inevitable.  This can be
quantified by observing that the particles have a slow net drift.
That is, $\langle \Delta \vec{r} \rangle = \vec{v}(t) \tau$ with
a slowly varying velocity $\vec{v}$, for small lag times $\tau$,
and where the angle brackets indicate an average over the
displacements $\Delta \vec{r}$ of all particles at a given time.
By examining the two-particle correlation functions
(Eqns.~\ref{eq:d2p}), it can be seen that such a drift will not
affect the antiparallel correlations, but will increase the
measured value for the parallel correlation functions $D_{x+}$
and $D_{y+}$.  If $\vec{v}$ was time-independent, then such drift
is straightforward to detect and remove from the particle
trajectories.  However, we often find that $\vec{v}(t)$ has a
slow but nontrivial time dependence, and this then makes its
influence on $D_{x+}$ and $D_{y+}$ depend on $\tau$.  Moreover,
due to hydrodynamic interactions, particle motions should have
long-range correlations even in the absence of flow, so there is
a danger that by removing correlated motion of all the particles,
some of the signal from hydrodynamic correlations is lost.
As already mentioned, the negative correlations are not affected
at all, as they measure the relative displacement of particles,
and any center of mass motion cancels out.  Likewise, single
particle measurements of $\langle \Delta r^2 \rangle$ are barely
affected by slow drift; to be safe, we calculate our single
particle data using $\langle (\Delta \vec{r} - \langle \Delta
\vec{r} \rangle)^2 \rangle = \langle \Delta r^2 \rangle - \langle
(\Delta r)^2 \rangle$.

To deal with the effects of drift on $D_{x+}$ and $D_{y+}$,
we compute $\langle \Delta \vec{r}(t) \rangle$ at every time $t$
using displacements with lag time $1/30$~s, the time between
images, as discussed in Sec.~\ref{sec:exp}.  We then integrate
$\langle \Delta \vec{r}(t) \rangle$ to get a trajectory of the
center of mass of all of the particles, $\vec{r}(t)$.  Next, we
smooth this with a running average over $T$ time steps.  We then
subtract the smoothed $\vec{r}(t)$ from each individual particle
trajectory, to bring the individual particle trajectories into
the moving reference frame.  In some
cases we do not do this smoothing, corresponding to the $T=1$
limit where the center of mass is forced to be motionless once
the trajectories are brought into the moving reference frame.

Figure \ref{fig:smoo} shows our analysis of how $T$ affects
two particle correlations for the same soap film as in
Fig.~\ref{fig:Dxy}, and for several choices of $\tau$.  All values
of the smoothing parameter $T$ and lag time $\tau$ are given in
terms of the frame rate of the camera, so $\tau=1$ corresponds to
1/30~s for example.  We desire that our results should be $\tau$
independent ideally.  Indeed, as should be, the $D_{x-}/\tau$
data all collapse for all smoothing parameters $T$ and lag times
$\tau$, as mathematically our procedure leaves the antiparallel
correlations unchanged.  These are the lower curves in
Fig.~\ref{fig:smoo} that increase for $R > 2$~$\mu$m.

For $T = 1$ in Fig.~\ref{fig:smoo}(a), the positive correlations
$D_{x+}$ have very slight $\tau$ dependency, for lower $\tau$. This
is likely due to artificial subtraction of positive correlations
as explained above.  For $T=8$ in Fig.~\ref{fig:smoo}(b), the
positive correlations at higher $\tau = 8, 12, 16$ collapse, but
the positive correlations curve for $\tau = 4$ is much higher. This
is due to lack of drift subtraction for $\tau < 8$ when $T = 8$.
This is evidence that $\vec{v}(t)$ changes even on a fairly
quick time scale of $\tau=4$ (corresponding to 4/30~s).

Figure~\ref{fig:smoo}(c) looks at different smoothing parameters
$T = 1, 2, 4, 8$, for the same lag time $\tau = 4$. The curves
do not collapse on each other for reasons explained above.
Fig.~\ref{fig:smoo}(d) looks at the same as (c), but for $\tau =
12$. The curves for different $T$ collapse nicely on top of each
other for $\tau = 12$. Hence, at higher lag times $\tau$s, we are
confident that smoothing does not affect our results, as long as
the smoothing parameter $T$ is chosen to be shorter than $\tau$.
In order for uniform treatment of all samples, we analyze
all movies using $T=1$.  For each movie,
we average the data over a wide range of $\tau$'s.  The smallest
$\tau$ is always 2 video frames (66~ms).  The largest $\tau$ is
chosen for each individual movie to be the largest one for which
data are available, in other words, the largest duration over which
individual particles are tracked; this is at most 1~s.  We
compute Eqns.~\ref{eq:d2p} for each $\tau$, fit to find the
parameters $A$, $B$, and $L$, and then average those parameters
over the different $\tau$'s.  The standard deviations of those
data lead to the uncertainties shown in Fig.~\ref{fig:ABLvshd4}.
While the evidence of Fig.~\ref{fig:smoo} gives support for our
choice $T=1$, we note that this must remove some real correlated
motion and may result in a lower measured average value of $A$.
$A$ depends on the intersection of the $T$-dependent parallel
correlations with the $T$-independent antiparallel correlations.
As noted in Sec.~\ref{Subsec:twopart}, we find $A/D_{1p}=1.83 \pm
0.09$ rather than the expected result $A/D_{1p}=2$.  However, we
find that using larger values for $T$ only gives more uncertain
results for the data shown in Fig.~\ref{fig:ABLvshd4}.



\begin{thebibliography}{36}
\expandafter\ifx\csname natexlab\endcsname\relax\def\natexlab#1{#1}\fi
\expandafter\ifx\csname bibnamefont\endcsname\relax
  \def\bibnamefont#1{#1}\fi
\expandafter\ifx\csname bibfnamefont\endcsname\relax
  \def\bibfnamefont#1{#1}\fi
\expandafter\ifx\csname citenamefont\endcsname\relax
  \def\citenamefont#1{#1}\fi
\expandafter\ifx\csname url\endcsname\relax
  \def\url#1{\texttt{#1}}\fi
\expandafter\ifx\csname urlprefix\endcsname\relax\def\urlprefix{URL }\fi
\providecommand{\bibinfo}[2]{#2}
\providecommand{\eprint}[2][]{\url{#2}}

\bibitem[{\citenamefont{Couder et~al.}(1989)\citenamefont{Couder, Chomaz, and
  Rabaud}}]{hydrodynamics}
\bibinfo{author}{\bibfnamefont{Y.}~\bibnamefont{Couder}},
  \bibinfo{author}{\bibfnamefont{J.~M.} \bibnamefont{Chomaz}},
  \bibnamefont{and} \bibinfo{author}{\bibfnamefont{M.}~\bibnamefont{Rabaud}},
  \bibinfo{journal}{Physica D: Nonlinear Phenomena}
  \textbf{\bibinfo{volume}{37}}, \bibinfo{pages}{384} (\bibinfo{year}{1989}).

\bibitem[{\citenamefont{Plateau}(1873)}]{plateau}
\bibinfo{author}{\bibfnamefont{J.}~\bibnamefont{Plateau}},
  \emph{\bibinfo{title}{Statique Exp\'{e}rimentale et Th\'{e}orique des
  Liquides Soumis aux Seules Forces Mol\'{e}culaires}}
  (\bibinfo{publisher}{Gauthier-Villars, Paris}, \bibinfo{year}{1873}).

\bibitem[{\citenamefont{Gibbs}(1931)}]{gibbs}
\bibinfo{author}{\bibfnamefont{J.~W.} \bibnamefont{Gibbs}},
  \emph{\bibinfo{title}{The Collected Works}} (\bibinfo{publisher}{Longmans
  Green, New York}, \bibinfo{year}{1931}).

\bibitem[{\citenamefont{Prasad and Weeks}(2009{\natexlab{a}})}]{vikPRE}
\bibinfo{author}{\bibfnamefont{V.}~\bibnamefont{Prasad}} \bibnamefont{and}
  \bibinfo{author}{\bibfnamefont{E.~R.} \bibnamefont{Weeks}},
  \bibinfo{journal}{Phys. Rev. E} \textbf{\bibinfo{volume}{80}},
  \bibinfo{pages}{026309} (\bibinfo{year}{2009}{\natexlab{a}}).

\bibitem[{\citenamefont{Prasad and Weeks}(2009{\natexlab{b}})}]{vikPRL}
\bibinfo{author}{\bibfnamefont{V.}~\bibnamefont{Prasad}} \bibnamefont{and}
  \bibinfo{author}{\bibfnamefont{E.~R.} \bibnamefont{Weeks}},
  \bibinfo{journal}{Phys. Rev. Lett.} \textbf{\bibinfo{volume}{102}},
  \bibinfo{pages}{178302} (\bibinfo{year}{2009}{\natexlab{b}}).

\bibitem[{\citenamefont{Cheung et~al.}(1996)\citenamefont{Cheung, Hwang, Wu,
  and Choi}}]{singlesoapfilm}
\bibinfo{author}{\bibfnamefont{C.}~\bibnamefont{Cheung}},
  \bibinfo{author}{\bibfnamefont{Y.~H.} \bibnamefont{Hwang}},
  \bibinfo{author}{\bibfnamefont{X.-l.} \bibnamefont{Wu}}, \bibnamefont{and}
  \bibinfo{author}{\bibfnamefont{H.~J.} \bibnamefont{Choi}},
  \bibinfo{journal}{Phys. Rev. Lett.} \textbf{\bibinfo{volume}{76}},
  \bibinfo{pages}{2531} (\bibinfo{year}{1996}).

\bibitem[{\citenamefont{Zhang et~al.}(2000)\citenamefont{Zhang, Childress,
  Libchaber, and Shelley}}]{1dflags}
\bibinfo{author}{\bibfnamefont{J.}~\bibnamefont{Zhang}},
  \bibinfo{author}{\bibfnamefont{S.}~\bibnamefont{Childress}},
  \bibinfo{author}{\bibfnamefont{A.}~\bibnamefont{Libchaber}},
  \bibnamefont{and} \bibinfo{author}{\bibfnamefont{M.}~\bibnamefont{Shelley}},
  \bibinfo{journal}{Nature} \textbf{\bibinfo{volume}{408}},
  \bibinfo{pages}{835} (\bibinfo{year}{2000}).

\bibitem[{\citenamefont{Rutgers}(1998)}]{rutgers98}
\bibinfo{author}{\bibfnamefont{M.~A.} \bibnamefont{Rutgers}},
  \bibinfo{journal}{Phys. Rev. Lett.} \textbf{\bibinfo{volume}{81}},
  \bibinfo{pages}{2244} (\bibinfo{year}{1998}).

\bibitem[{\citenamefont{Tien and Ottova}(2001)}]{soaplipid}
\bibinfo{author}{\bibfnamefont{H.~T.} \bibnamefont{Tien}} \bibnamefont{and}
  \bibinfo{author}{\bibfnamefont{A.~L.} \bibnamefont{Ottova}},
  \bibinfo{journal}{Journal of Membrane Science}
  \textbf{\bibinfo{volume}{189}}, \bibinfo{pages}{83} (\bibinfo{year}{2001}).

\bibitem[{\citenamefont{Martin et~al.}(1998)\citenamefont{Martin, Wu, Goldburg,
  and Rutgers}}]{rutgers98a}
\bibinfo{author}{\bibfnamefont{B.~K.} \bibnamefont{Martin}},
  \bibinfo{author}{\bibfnamefont{X.~L.} \bibnamefont{Wu}},
  \bibinfo{author}{\bibfnamefont{W.~I.} \bibnamefont{Goldburg}},
  \bibnamefont{and} \bibinfo{author}{\bibfnamefont{M.~A.}
  \bibnamefont{Rutgers}}, \bibinfo{journal}{Phys. Rev. Lett.}
  \textbf{\bibinfo{volume}{80}}, \bibinfo{pages}{3964} (\bibinfo{year}{1998}).

\bibitem[{\citenamefont{Burgess et~al.}(1999)\citenamefont{Burgess, Bizon,
  McCormick, Swift, and Swinney}}]{burgess99}
\bibinfo{author}{\bibfnamefont{J.~M.} \bibnamefont{Burgess}},
  \bibinfo{author}{\bibfnamefont{C.}~\bibnamefont{Bizon}},
  \bibinfo{author}{\bibfnamefont{W.~D.} \bibnamefont{McCormick}},
  \bibinfo{author}{\bibfnamefont{J.~B.} \bibnamefont{Swift}}, \bibnamefont{and}
  \bibinfo{author}{\bibfnamefont{H.~L.} \bibnamefont{Swinney}},
  \bibinfo{journal}{Phys. Rev. E} \textbf{\bibinfo{volume}{60}},
  \bibinfo{pages}{715} (\bibinfo{year}{1999}).

\bibitem[{\citenamefont{Trapeznikov}(1957)}]{trapeznikov}
\bibinfo{author}{\bibfnamefont{A.~A.} \bibnamefont{Trapeznikov}},
  \bibinfo{journal}{Proceedings of the 2nd International Congress on Surface
  Activity} pp. \bibinfo{pages}{242--258} (\bibinfo{year}{1957}).

\bibitem[{\citenamefont{Saffman and Delbr\"{u}ck}(1975)}]{sd}
\bibinfo{author}{\bibfnamefont{P.~G.} \bibnamefont{Saffman}} \bibnamefont{and}
  \bibinfo{author}{\bibfnamefont{M.}~\bibnamefont{Delbr\"{u}ck}},
  \bibinfo{journal}{Proc. Nat. Acad. Sci.} \textbf{\bibinfo{volume}{72}},
  \bibinfo{pages}{3111} (\bibinfo{year}{1975}).

\bibitem[{\citenamefont{Stone and Armand}(1998)}]{hydrodynamicssd}
\bibinfo{author}{\bibfnamefont{H.~A.} \bibnamefont{Stone}} \bibnamefont{and}
  \bibinfo{author}{\bibfnamefont{A.}~\bibnamefont{Armand}},
  \bibinfo{journal}{J. Fluid Mech.} \textbf{\bibinfo{volume}{369}},
  \bibinfo{pages}{151} (\bibinfo{year}{1998}).

\bibitem[{\citenamefont{Stone}(2010)}]{stoneprize}
\bibinfo{author}{\bibfnamefont{H.~A.} \bibnamefont{Stone}},
  \bibinfo{journal}{J. Fluid Mech.} \textbf{\bibinfo{volume}{645}},
  \bibinfo{pages}{1} (\bibinfo{year}{2010}).

\bibitem[{\citenamefont{Eremin et~al.}(2011)\citenamefont{Eremin, Baumgarten,
  Harth, Stannarius, Nguyen, Goldfain, Park, Maclennan, Glaser, and
  Clark}}]{eremin11}
\bibinfo{author}{\bibfnamefont{A.}~\bibnamefont{Eremin}},
  \bibinfo{author}{\bibfnamefont{S.}~\bibnamefont{Baumgarten}},
  \bibinfo{author}{\bibfnamefont{K.}~\bibnamefont{Harth}},
  \bibinfo{author}{\bibfnamefont{R.}~\bibnamefont{Stannarius}},
  \bibinfo{author}{\bibfnamefont{Z.~H.} \bibnamefont{Nguyen}},
  \bibinfo{author}{\bibfnamefont{A.}~\bibnamefont{Goldfain}},
  \bibinfo{author}{\bibfnamefont{C.~S.} \bibnamefont{Park}},
  \bibinfo{author}{\bibfnamefont{J.~E.} \bibnamefont{Maclennan}},
  \bibinfo{author}{\bibfnamefont{M.~A.} \bibnamefont{Glaser}},
  \bibnamefont{and} \bibinfo{author}{\bibfnamefont{N.~A.} \bibnamefont{Clark}},
  \bibinfo{journal}{Phys. Rev. Lett.} \textbf{\bibinfo{volume}{107}},
  \bibinfo{pages}{268301} (\bibinfo{year}{2011}).

\bibitem[{\citenamefont{Domanov et~al.}(2011)\citenamefont{Domanov, Aimon,
  Toombes, Renner, Quemeneur, Triller, Turner, and Bassereau}}]{domanov2011}
\bibinfo{author}{\bibfnamefont{Y.~A.} \bibnamefont{Domanov}},
  \bibinfo{author}{\bibfnamefont{S.}~\bibnamefont{Aimon}},
  \bibinfo{author}{\bibfnamefont{G.~E.~S.} \bibnamefont{Toombes}},
  \bibinfo{author}{\bibfnamefont{M.}~\bibnamefont{Renner}},
  \bibinfo{author}{\bibfnamefont{F.}~\bibnamefont{Quemeneur}},
  \bibinfo{author}{\bibfnamefont{A.}~\bibnamefont{Triller}},
  \bibinfo{author}{\bibfnamefont{M.~S.} \bibnamefont{Turner}},
  \bibnamefont{and}
  \bibinfo{author}{\bibfnamefont{P.}~\bibnamefont{Bassereau}},
  \bibinfo{journal}{Proc. Nat. Acad. Sci.} \textbf{\bibinfo{volume}{108}},
  \bibinfo{pages}{12605} (\bibinfo{year}{2011}).

\bibitem[{\citenamefont{Einstein}(1905)}]{einstein1905}
\bibinfo{author}{\bibfnamefont{A.}~\bibnamefont{Einstein}},
  \bibinfo{journal}{Ann. Phys.} \textbf{\bibinfo{volume}{322}},
  \bibinfo{pages}{549} (\bibinfo{year}{1905}).

\bibitem[{\citenamefont{Sutherland}(1905)}]{sutherland1905}
\bibinfo{author}{\bibfnamefont{W.}~\bibnamefont{Sutherland}},
  \bibinfo{journal}{Phil. Mag. Series 6} \textbf{\bibinfo{volume}{9}},
  \bibinfo{pages}{781} (\bibinfo{year}{1905}).

\bibitem[{\citenamefont{Hughes et~al.}(1981)\citenamefont{Hughes, Pailthorpe,
  and White}}]{hpw}
\bibinfo{author}{\bibfnamefont{B.~D.} \bibnamefont{Hughes}},
  \bibinfo{author}{\bibfnamefont{B.~A.} \bibnamefont{Pailthorpe}},
  \bibnamefont{and} \bibinfo{author}{\bibfnamefont{L.~R.} \bibnamefont{White}},
  \bibinfo{journal}{J. Fluid Mech.} \textbf{\bibinfo{volume}{110}},
  \bibinfo{pages}{349} (\bibinfo{year}{1981}).

\bibitem[{\citenamefont{Saffman}(1976)}]{saffman1976}
\bibinfo{author}{\bibfnamefont{P.~G.} \bibnamefont{Saffman}},
  \bibinfo{journal}{J. Fluid Mech.} \textbf{\bibinfo{volume}{73}},
  \bibinfo{pages}{593} (\bibinfo{year}{1976}).

\bibitem[{\citenamefont{Sickert et~al.}(2007)\citenamefont{Sickert, Rondelez,
  and Stone}}]{Langmuirmonolayer}
\bibinfo{author}{\bibfnamefont{M.}~\bibnamefont{Sickert}},
  \bibinfo{author}{\bibfnamefont{F.}~\bibnamefont{Rondelez}}, \bibnamefont{and}
  \bibinfo{author}{\bibfnamefont{H.~A.} \bibnamefont{Stone}},
  \bibinfo{journal}{Europhys. Lett.} \textbf{\bibinfo{volume}{79}},
  \bibinfo{pages}{66005} (\bibinfo{year}{2007}).

\bibitem[{\citenamefont{Petrov and Schwille}(2008)}]{petrov08}
\bibinfo{author}{\bibfnamefont{E.~P.} \bibnamefont{Petrov}} \bibnamefont{and}
  \bibinfo{author}{\bibfnamefont{P.}~\bibnamefont{Schwille}},
  \bibinfo{journal}{Biophys. J.} \textbf{\bibinfo{volume}{94}},
  \bibinfo{pages}{L41} (\bibinfo{year}{2008}).

\bibitem[{\citenamefont{Petrov et~al.}(2012)\citenamefont{Petrov, Petrosyan,
  and Schwille}}]{petrov12}
\bibinfo{author}{\bibfnamefont{E.~P.} \bibnamefont{Petrov}},
  \bibinfo{author}{\bibfnamefont{R.}~\bibnamefont{Petrosyan}},
  \bibnamefont{and} \bibinfo{author}{\bibfnamefont{P.}~\bibnamefont{Schwille}},
  \bibinfo{journal}{Soft Matter} \textbf{\bibinfo{volume}{8}},
  \bibinfo{pages}{7552} (\bibinfo{year}{2012}).

\bibitem[{\citenamefont{Nguyen et~al.}(2010)\citenamefont{Nguyen, Atkinson,
  Park, Maclennan, Glaser, and Clark}}]{sdliqfilm}
\bibinfo{author}{\bibfnamefont{Z.~H.} \bibnamefont{Nguyen}},
  \bibinfo{author}{\bibfnamefont{M.}~\bibnamefont{Atkinson}},
  \bibinfo{author}{\bibfnamefont{C.~S.} \bibnamefont{Park}},
  \bibinfo{author}{\bibfnamefont{J.}~\bibnamefont{Maclennan}},
  \bibinfo{author}{\bibfnamefont{M.}~\bibnamefont{Glaser}}, \bibnamefont{and}
  \bibinfo{author}{\bibfnamefont{N.}~\bibnamefont{Clark}},
  \bibinfo{journal}{Phys. Rev. Lett.} \textbf{\bibinfo{volume}{105}},
  \bibinfo{pages}{268304} (\bibinfo{year}{2010}).

\bibitem[{\citenamefont{Crocker et~al.}(2000)\citenamefont{Crocker, Valentine,
  Weeks, Gisler, Kaplan, Yodh, and Weitz}}]{two-pointmicrorh}
\bibinfo{author}{\bibfnamefont{J.~C.}~\bibnamefont{Crocker}},
  \bibinfo{author}{\bibfnamefont{M.~T.} \bibnamefont{Valentine}},
  \bibinfo{author}{\bibfnamefont{E.~R.} \bibnamefont{Weeks}},
  \bibinfo{author}{\bibfnamefont{T.}~\bibnamefont{Gisler}},
  \bibinfo{author}{\bibfnamefont{P.~D.} \bibnamefont{Kaplan}},
  \bibinfo{author}{\bibfnamefont{A.~G.} \bibnamefont{Yodh}}, \bibnamefont{and}
  \bibinfo{author}{\bibfnamefont{D.~A.} \bibnamefont{Weitz}},
  \bibinfo{journal}{Phys. Rev. Lett.} \textbf{\bibinfo{volume}{85}},
  \bibinfo{pages}{888} (\bibinfo{year}{2000}).

\bibitem[{\citenamefont{Levine and Lubensky}(2000)}]{two-pointmicrorhtheory}
\bibinfo{author}{\bibfnamefont{A.~J.} \bibnamefont{Levine}} \bibnamefont{and}
  \bibinfo{author}{\bibfnamefont{T.~C.} \bibnamefont{Lubensky}},
  \bibinfo{journal}{Phys. Rev. Lett.} \textbf{\bibinfo{volume}{85}},
  \bibinfo{pages}{1774} (\bibinfo{year}{2000}).

\bibitem[{\citenamefont{Di~Leonardo
  et~al.}(2008{\natexlab{a}})\citenamefont{Di~Leonardo, Keen, Ianni, Leach,
  Padgett, and Ruocco}}]{2Dleonardo}
\bibinfo{author}{\bibfnamefont{R.}~\bibnamefont{Di~Leonardo}},
  \bibinfo{author}{\bibfnamefont{S.}~\bibnamefont{Keen}},
  \bibinfo{author}{\bibfnamefont{F.}~\bibnamefont{Ianni}},
  \bibinfo{author}{\bibfnamefont{J.}~\bibnamefont{Leach}},
  \bibinfo{author}{\bibfnamefont{M.~J.} \bibnamefont{Padgett}},
  \bibnamefont{and} \bibinfo{author}{\bibfnamefont{G.}~\bibnamefont{Ruocco}},
  \bibinfo{journal}{Phys. Rev. E} \textbf{\bibinfo{volume}{78}},
  \bibinfo{pages}{031406} (\bibinfo{year}{2008}{\natexlab{a}}).

\bibitem[{\citenamefont{Fischer et~al.}(2006)\citenamefont{Fischer, Dhar, and
  Heinig}}]{spheredrag}
\bibinfo{author}{\bibfnamefont{T.~M.} \bibnamefont{Fischer}},
  \bibinfo{author}{\bibfnamefont{P.}~\bibnamefont{Dhar}}, \bibnamefont{and}
  \bibinfo{author}{\bibfnamefont{P.}~\bibnamefont{Heinig}},
  \bibinfo{journal}{J. Fluid Mech.} \textbf{\bibinfo{volume}{558}},
  \bibinfo{pages}{451} (\bibinfo{year}{2006}).

\bibitem[{\citenamefont{Crocker and Grier}(1996)}]{idlref}
\bibinfo{author}{\bibfnamefont{J.~C.} \bibnamefont{Crocker}} \bibnamefont{and}
  \bibinfo{author}{\bibfnamefont{D.~G.} \bibnamefont{Grier}},
  \bibinfo{journal}{J. Colloid Interface Sci.} \textbf{\bibinfo{volume}{179}},
  \bibinfo{pages}{298} (\bibinfo{year}{1996}).

\bibitem[{\citenamefont{Di~Leonardo
  et~al.}(2008{\natexlab{b}})\citenamefont{Di~Leonardo, Saglimbeni, and
  Ruocco}}]{dileonardo08}
\bibinfo{author}{\bibfnamefont{R.}~\bibnamefont{Di~Leonardo}},
  \bibinfo{author}{\bibfnamefont{F.}~\bibnamefont{Saglimbeni}},
  \bibnamefont{and} \bibinfo{author}{\bibfnamefont{G.}~\bibnamefont{Ruocco}},
  \bibinfo{journal}{Phys. Rev. Lett.} \textbf{\bibinfo{volume}{100}},
  \bibinfo{pages}{106103} (\bibinfo{year}{2008}{\natexlab{b}}).

\bibitem[{\citenamefont{Kaz et~al.}(2012)\citenamefont{Kaz, McGorty, Mani,
  Brenner, and Manoharan}}]{kaz12}
\bibinfo{author}{\bibfnamefont{D.~M.} \bibnamefont{Kaz}},
  \bibinfo{author}{\bibfnamefont{R.}~\bibnamefont{McGorty}},
  \bibinfo{author}{\bibfnamefont{M.}~\bibnamefont{Mani}},
  \bibinfo{author}{\bibfnamefont{M.~P.} \bibnamefont{Brenner}},
  \bibnamefont{and} \bibinfo{author}{\bibfnamefont{V.~N.}
  \bibnamefont{Manoharan}}, \bibinfo{journal}{Nat. Mater.}
  \textbf{\bibinfo{volume}{11}}, \bibinfo{pages}{138} (\bibinfo{year}{2012}).

\bibitem[{\citenamefont{Wu et~al.}(2001)\citenamefont{Wu, Levine, Rutgers,
  Kellay, and Goldburg}}]{sfinfrared}
\bibinfo{author}{\bibfnamefont{X.~L.} \bibnamefont{Wu}},
  \bibinfo{author}{\bibfnamefont{R.}~\bibnamefont{Levine}},
  \bibinfo{author}{\bibfnamefont{M.}~\bibnamefont{Rutgers}},
  \bibinfo{author}{\bibfnamefont{H.}~\bibnamefont{Kellay}}, \bibnamefont{and}
  \bibinfo{author}{\bibfnamefont{W.~I.} \bibnamefont{Goldburg}},
  \bibinfo{journal}{Rev. Sci. Inst.} \textbf{\bibinfo{volume}{72}},
  \bibinfo{pages}{2467} (\bibinfo{year}{2001}).

\bibitem[{\citenamefont{Bechhoefer et~al.}(1997)\citenamefont{Bechhoefer,
  G\'{e}minard, Bocquet, and Oswald}}]{diffusionliqfillm}
\bibinfo{author}{\bibfnamefont{J.}~\bibnamefont{Bechhoefer}},
  \bibinfo{author}{\bibfnamefont{J.~C.} \bibnamefont{G\'{e}minard}},
  \bibinfo{author}{\bibfnamefont{L.}~\bibnamefont{Bocquet}}, \bibnamefont{and}
  \bibinfo{author}{\bibfnamefont{P.}~\bibnamefont{Oswald}},
  \bibinfo{journal}{Phys. Rev. Lett.} \textbf{\bibinfo{volume}{79}},
  \bibinfo{pages}{4922} (\bibinfo{year}{1997}).

\bibitem[{\citenamefont{Sur and Kyu~Pak}(2001)}]{capillarycolloidfilm}
\bibinfo{author}{\bibfnamefont{J.}~\bibnamefont{Sur}} \bibnamefont{and}
  \bibinfo{author}{\bibfnamefont{H.~K.}~\bibnamefont{Pak}},
  \bibinfo{journal}{Phys. Rev. Lett.} \textbf{\bibinfo{volume}{86}},
  \bibinfo{pages}{4326} (\bibinfo{year}{2001}).

\bibitem[{\citenamefont{Nikolaides et~al.}(2002)\citenamefont{Nikolaides,
  Bausch, Hsu, Dinsmore, Brenner, Gay, and Weitz}}]{capillaryinterface}
\bibinfo{author}{\bibfnamefont{M.~G.} \bibnamefont{Nikolaides}},
  \bibinfo{author}{\bibfnamefont{A.~R.} \bibnamefont{Bausch}},
  \bibinfo{author}{\bibfnamefont{M.~F.} \bibnamefont{Hsu}},
  \bibinfo{author}{\bibfnamefont{A.~D.} \bibnamefont{Dinsmore}},
  \bibinfo{author}{\bibfnamefont{M.~P.} \bibnamefont{Brenner}},
  \bibinfo{author}{\bibfnamefont{C.}~\bibnamefont{Gay}}, \bibnamefont{and}
  \bibinfo{author}{\bibfnamefont{D.~A.} \bibnamefont{Weitz}},
  \bibinfo{journal}{Nature} \textbf{\bibinfo{volume}{420}},
  \bibinfo{pages}{299} (\bibinfo{year}{2002}).

\end{thebibliography}
\end{document}